# The Role of Life Cycle and Migration in Selection for Variance in Offspring Number

Max Shpak <mshpak@tiem.utk.edu>, Department of Ecology and Evolutionary Biol‒ogy, University of Tennessee. Knoxville, TN 37996 USA, (865) 974–4605, (865) 974–3067 (fax)

Stephen Proulx <proulx@proulxresearch.org>, Department of Ecology and Evolution‒ary Biology, Iowa State University. Ames, IA 50011 USA, (515) 294–0272

## Abstract

For two genotypes that have the same mean number of offspring but differ in the variance in offspring number, natural selection will favor the genotype with lower variance. The concept of fitness becomes cloudy under these conditions because the outcome of evolution is not deterministic. However, the effect of variance in offspring number on the fixation probability of mutant strategies has been calculated under several scenarios with the general conclusion that variance in offspring number reduces fitness but only in proportion to the inverse of the population size ( Gillespie 1974, Proulx 2000). This relationship becomes more complicated under a metapopulation scenario where the "effective" population size depends on migration rate, population structure, and life cycle. We show that under hard selection and weak migration fitness in a metapopulation composed of equal sized demes is determined by deme size. Conversely, for high migration rates and hard selection the effective fitness depends on the total size of the metapopulation. Interestingly, under soft selection there is no effect of migration or neighboring population structure on effective fitness, and fitness depends only on deme size. We use individual based simulations in developed in Shpak (2005) to validate our analytical approximations and investigate deviations of our assumption of equal deme size.

## Introduction: A Summary of Previous Work

Consider two competing alleles in a population. An individual carrying one allele or the complementary copy produces an average of $\mu_1$ and $\mu_2$ offspring in each generation (respectively), with corresponding variance in offspring number per clutch of $\sigma_1^2$, $\sigma_2^2$. It has been shown (Gillespie 1974) that the fixation probabilities cannot be predicted from arithmetic mean fitness alone, i.e. for sufficiently high variance $\sigma_2^2 < \sigma_1^2$, the second allele can have a higher probability of fixation even when $\mu_2 < \mu_1$. The condition for the first strategy having a higher probability of fixation than the second first derived by Gillespie is

$$(1) \qquad \mu_2 - \frac{\sigma_2^2}{n} \;<\; \mu_1 - \frac{\sigma_1^2}{n}$$

This inequality states that a higher mean fitness is not necessarily favored by natural selection if the variance is high, because of the possibility of the high mean, high variance strategy producing a lower number of offspring than the competitor in any given trial. It should be intuitively obvious that this effect is less pronounced in larger populations due to the averaging effects of reduced sample variance for large n. This has important implications for organisms where there is a trade-off between producing a high average number of offspring and reducing the variance, as the outcome of selection for one strategy or another will depend on the population size. For example, a semelparous reproductive strategy, all else being equal, has a higher variance in surviving offspring when clutches succeed or fail as a whole than an iteroparous strategy with the same mean number of offspring. If the semelparous strategy also has a somewhat higher mean, whether it or a competing iteroparous strategy becomes fixed will depend on whether the population is large or small.

Gillespie calculated the relationship between mean, variance, and "effective" fitness by a change of variables in a diffusion equation and by collecting the coefficients associated with the first derivative of the density function and frequency p. Following Proulx (2000), the relationship between variance in offspring number and selection can also be derived from first principles by calculating the expected change in the number of individuals $x_1(t)$ carrying alleles of the first type, with

$$(2) \quad x_1(t+1) = \sum_{i=1}^{x_1} (\mu_1 + \xi_1[i]),$$

where $\xi_1[i]$ is a random variable with mean 0 and variance $\sigma_1^2$ (with corresponding equations for the number of individuals with the second allele $x_2$). Note that the only contribution to variance considered here is that due to variation in offspring number, the variance due to genetic drift (sampling of a fixed number of individuals from an offspring pool) is not explicitly considered as contributing to $\xi$.

The expected frequency in the next time step is:

$$(3) \quad p(t+1) = \frac{\sum_{i=1}^{x_1} (\mu_1 + \xi_1[i])}{\sum_{i=1}^{x_1} (\mu_1 + \xi_1[i]) + \sum_{i=1}^{x_2} (\mu_2 + \xi_2[i])}.$$

At this stage, the effects of drift can be included by considering another sampling process that introduces variance. However, so long as the sampling process is fair, the expected frequency of allele 1 in the next generation will be given by (3) (Proulx 2000). In order to accurately describe selection in terms of first and second order terms alone, it must be assumed that $\xi_1[i]$ is small of the order $\epsilon$. For convenience, the random variable $\xi_i$ is replaced by $\epsilon z[i]$, where $z[i]$ is a rescaled random variable of order ~1 and $\epsilon$ is a constant <<1. Equation (3) then becomes

(4) $\quad p(t+1) =$
$$\frac{x_1 \mu_1 + \sum_{i=1}^{x_1} \epsilon z_1[i]}{x_1 \mu_1 + \sum_{i=1}^{x_1} \epsilon z_1[i] + x_2 \mu_2 + \sum_{i=1}^{x_2} \epsilon z_2[i]}.$$

The above expression can be written as a Taylor expansion (up to second order terms) as a function of $\epsilon$ about $\epsilon_0 = 0$:

(5)
$$\frac{x_1 \mu_1}{x_1 \mu_1 + x_2 \mu_2} + \epsilon \left( \left( (x_1 \mu_1 + x_2 \mu_2) \sum_{i=1}^{x_1} z_1[i] - x_1 \mu_1 \left( \sum_{i=1}^{x_1} z_1[i] + \sum_{i=1}^{x_2} z_2[i] \right) \right) \Big/ (x_1 \mu_1 + x_2 \mu_2)^2 \right) -$$
$$\epsilon^2 \left( \frac{\sum_{i=1}^{x_1} z_1[i] \left( \sum_{i=1}^{x_1} z_1[i] + \sum_{i=1}^{x_2} z_2[i] \right)}{(x_1 \mu_1 + x_2 \mu_2)^2} + \frac{x_1 \mu_1 \left( \sum_{i=1}^{x_1} z_1[i] + \sum_{i=1}^{x_2} z_2[i] \right)^2}{(x_1 \mu_1 + x_2 \mu_2)^3} \right).$$

The expansion can be simplified due to the fact that each reproductive event in a population is independent, so that the covariances cov($z_1$[i],$z_1$[j])=0 and cov($z_1$[i],$z_2$[j])=0 for all i,j. Furthermore, the expectation values of $z_1$, $z_2$ are 0, so that the expected means and variances are:

$$E\left[ \sum_{i=1}^{x_1} z_1[i] \right] = E\left[ \sum_{i=1}^{x_2} z_2[i] \right] = 0$$

$$E\left[ \left( \sum_{i=1}^{x_1} z_1[i] \right)^2 \right] = x_1 s_1 \; ; \; E\left[ \left( \sum_{i=1}^{x_2} z_2[i] \right)^2 \right] = x_2 s_2$$

The terms $s_1, s_2$ are rescaled variance terms such that $\sigma_i^2 = \epsilon^2 s_i$. Substituting these variance and covariance relations into (5), together with the first moments $\mu$, one obtains the expectation value

$$(6) \quad E[p(t+1)] = \frac{x_1 \mu_1}{x_1 \mu_1 + x_2 \mu_2} - \frac{x_1 \sigma_1^2}{(x_1 \mu_1 + x_2 \mu_2)^2} + \frac{x_1 \mu_1 (x_1 \sigma_1^2 + x_2 \sigma_2^2)}{(x_1 \mu_1 + x_2 \mu_2)^3}$$

By abuse of notation, p(t)=p, the initial frequency of allele 1. Assuming fixed population size and soft selection, the final form of the expectation values is given substituting np for $x_1$ and n(1−p) for $x_2 = n - x_1$ (the factor of n in the numerator and denominator cancels):

$$(7) \quad E[p(t+1)] = \frac{p \mu_1}{p \mu_1 + (1-p) \mu_2} - \frac{(1-p) p (\mu_2 \sigma_1^2 - \mu_1 \sigma_2^2)}{n (p \mu_1 + (1-p) \mu_2)^3}$$

The mean change in allele frequency is just the difference between E[p(t+1)] and p, which gives

$$(8) \quad M(p) = E[p(t+1)] - p = \frac{p(1-p)(\mu_1 - \mu_2)}{p \mu_1 + (1-p) \mu_2} - \frac{(1-p) p (\mu_2 \sigma_1^2 - \mu_1 \sigma_2^2)}{n (p \mu_1 + (1-p) \mu_2)^3}$$

The first term in the expression represents the rate of change in allele frequency due to differences in mean fitness (the "deterministic" term) while the second term represents the selection differential on offspring variance.

In the limit where the variance values $\sigma^2$ are all small and the mean values $\mu$ are close to unity the above expression can be further simplified to the M(p) term of Gillespie (1974),

$$(9) \quad M(p) \approx p(1-p)\left((\mu_1 - \mu_2) - \frac{\sigma_1^2 - \sigma_2^2}{n}\right).$$

However, these conditions are seldom met in simulations (or in nature) so that equation (8) typically provides a better fit than equation (9) (Proulx 2000, and compare to Shpak 2005).

However, these conditions are seldom met in simulations (or in nature) so that equation (8) typically provides a better fit than equation (9) (Proulx 2000, and compare to Shpak 2005).

M(p) is the directional component in the diffusion approximation, i.e. the Kolmogorov backward equation (e.g. Kimura 1964, Crow and Kimura 1970, Ewens 1978)

$$(10) \quad \frac{\partial \phi}{\partial t} = M(p) \frac{\partial \phi}{\partial p} + \frac{V(p)}{2} \frac{\partial^2 \phi}{\partial p^2}$$

Unlike genetic drift, variance in offspring number within generations contributes both to the directional term M(p) and to the diffusion coefficient V(p), where V(p) is the variance in the change of allele frequency, i.e.

$$(11) \quad V(p) = E[(p(t+1) - p(t))^2] = E[p^2(t+1)] - 2 p E[p(t+1)] + p^2$$

The first quadratic term from the above is calculated

$$(12) \quad p^2(t+1) = \frac{\left(\sum_{i=1}^{x_1} (\mu_1 + \xi_1[i])\right)^2}{\left(\sum_{i=1}^{x_1} (\mu_1 + \xi_1[i]) + \sum_{i=1}^{x_2} (\mu_2 + \xi_2[i])\right)^2}$$

applying the same substitutions of variables and the same assumptions about the scaling of the mean and variance terms, the expectation of the second moment is:

$$(13) \quad E[p^2(t+1)] = \frac{n^2 p^2 \mu_1^2}{(n p \mu_1 + n (1-p) \mu_2)^2} - \frac{p (3 p \mu_1 - (1-p) \mu_2) \sigma_1^2}{n (p \mu_1 + (1-p) \mu_2)^3} + \frac{3 p^2 \mu_1^2 (p \sigma_1^2 + (1-p) \sigma_2^2)^2}{n (p \mu_1 + (1-p) \mu_2)^4}$$

combining with the known expression for E[p(t+1)] and simplifying, the variance in change of allele frequency is

$$(14) \quad V(p) = \frac{n^2 p^2 \mu_1^2}{(n p \mu_1 + n(1-p)\mu_2)^2} -$$

$$\frac{p(3p\mu_1 - (1-p)\mu_2)\sigma_1^2}{n(p\mu_1 + (1-p)\mu_2)^3} + \frac{3 p^2 \mu_1^2 (p\sigma_1^2 + (1-p)\sigma_2^2)^2}{n(p\mu_1 + (1-p)\mu_2)^4} -$$

$$2p\left(\frac{p\mu_1}{p\mu_1 + (1-p)\mu_2} - \frac{(1-p)p(\mu_2\sigma_1^2 - \mu_1\sigma_2^2)}{n(p\mu_1 + (1-p)\mu_2)^3}\right) + p^2$$

Again, with small variances and mean numbers of offspring near unity, this was shown in Proulx (2000) to simplify to the expression for V(p) in Gillespie (1974):

$$(15) \quad V(p) \approx \frac{p(1-p)}{2N}\left((1-p)\sigma_1^2 + p\sigma_2^2\right)$$

Given the first and second moments M(p), V(p), the Kolmogorov backward equation (10) can be solved for the fixation probability of an allele with initial frequency p (e.g. Kimura 1964),

$$(16)$$

$$U(p) = \frac{\int_0^p \text{Exp}\left[-\int 2\frac{M(x)}{V(x)}\,dx\right]dx}{\int_0^1 \text{Exp}\left[-\int 2\frac{M(x)}{V(x)}\,dx\right]dx} =$$

$$\frac{\int_0^p ((1-x)\sigma_1^2 + x\sigma_2^2)^{2\left(\frac{N(\mu_1-\mu_2)}{\sigma_1^2-\sigma_2^2}-1\right)}dx}{\int_0^1 ((1-x)\sigma_1^2 + x\sigma_2^2)^{2\left(\frac{N(\mu_1-\mu_2)}{\sigma_1^2-\sigma_2^2}-1\right)}dx}$$

The fixation probabilities of alleles that give different means and variances in offspring numbers have been analyzed for various parameters and found to be consistent with the fixation probabilities obtained from individual based simulation results in Shpak (2005).

The most significant results in the single deme model follow directly from (1), namely, that for nearly equal arithmetic mean numbers of offspring per generation, the

$\mu_1 \qquad \mu_2 \quad \sigma_2^2 \qquad \sigma_1^2$

$\hat{n}$

$\hat{n}$

genotype that produces the smaller variance in offspring number will be favored by selection. Furthermore, in the case where $\mu_1$ is larger than $\mu_2$ and $\sigma_2^2$ is smaller than $\sigma_1^2$, there will be a critical population size $\hat{n}$ at which the two alleles have equal fitness (and consequently, population sizes n<$\hat{n}$ will favor the lower mean, higher variance strategy versus larger population sizes $\hat{n}$<n corresponding to selection in favor of the first). This critical population size is calculated by setting both sides of (1) equal to one another:

$$(17) \qquad \hat{n} = \frac{\sigma_2^2 - \sigma_1^2}{\mu_1 - \mu_2}$$

This relationship between population size and clutch average/variance becomes less straightforward in the context of a metapopulation. If instead of a single, isolated deme, there are multiple demes exchanging migrants with one another, it seems apparent that the effective population size from the standpoint of selection on offspring variance will depend on deme structure and migration rate. For instance, in the absence of spatial structure, if there are D demes of n individuals, one would expect that for very low migration rates the effective fitness of a strategy would be approximated by $\mu - \frac{\sigma^2}{n}$ (nearly independent demes) while for very high migration rates (approaching complete mixing) the effective fitness would be closer to $\mu - \frac{\sigma^2}{nD}$ (complete mixing in the metapopulation), with values in the denominator between n and Dn for intermediate migration rates. This suggests that for values of $\mu$ and $\sigma^2$ such that n<$\hat{n}$ and $\hat{n}$<nD, there should also be a critical value of migration rate at which a high variance, lower mean strategy starts to be disfavored by selection.

This problem was investigated using individual based simulations in Shpak (2005). It was found that for the metapopulation scenario described above, with D demes with n individuals, each of which exchanges a fraction m/(D−1) with every neighbor, under certain cases there was indeed a critical value of m at which a strategy disfa–

$\hat{n}$

vored in a single deme of size $n<\hat{n}$ (i.e. higher mean, high variance) begins to be favored due to the effects of a greater "effective" population size under higher migration.

　　　　Significantly, it was found that migration only had this effect on selection for variance in a metapopulation if it occurred after reproduction but prior to selection. If reproduction and selection took place within individual demes and was followed by migration of post-selection adults, there was no effect and whatever strategy was favored in the individual demes in the absence of selection remained so even for high migration rates. The heuristic argument for why this is the case is that in the latter life cycle (Birth→Selection→Migration), the entire sampling process takes place within each deme, so that migration only accomplishes a "mixing" (and subsequent homogenization) of allele frequencies. In contrast, for the first life cycle (Birth→Migration→Selection), the pool of offspring that contribute to a given deme are sampled from the entire metapopulation, so that the effective sample variance depends on the contribution across demes as well as within demes in every generation.

　　　　Below, derivations of M(p) are presented for spatially unstructured metapopulations for two different life cycles. The derivations essentially follow the methods used in deriving equations (8-9) above (apart from the introduction of factors of (1-m) for "residents" and m for "migrants" for every deme and its neighbors). The expressions for the expected change in allele frequency are consistent with the numerical results and heuristic arguments in Shpak (2005), where it was argued that under certain life cycles (namely, Birth→Selection→Migration) the selection dynamics were essentially independent of both deme number and migration rate, while a life cycle where migration precedes selection gives the intuitive result where high migration rates and large numbers of demes reduce the effect of offspring variance.

## Expected Change in Allele Frequency in Metapopulations

The life cycle of an organism in a metapopulation can be broken down into the processes of migration, reproduction (birth), and natural selection. Here we assume that selection is assumed to be soft (e.g. Wallace 1970), so that the population size of any deme is held fixed at n. The fixed population size acts as a "normalization" in every generation. As in the single deme case, genetic drift is not explicitly considered (i.e. there is no variance contribution due to sampling of n offspring in each deme to main‐tained fixed population size). Reproduction is again described by every genotype in the ith deme producing $\mu+\xi_i=\mu+\epsilon z$ progeny, so that the random variable describing off‐spring number variance scales as $\epsilon \ll 1$.

A number of additional approximations are made in deriving the expected change in allele frequency in these model systems. For both the Birth→Selection→Migra‐tion (BMS) and Migration→Selection→Birth (MSB) life cycles, the contribution of migrants in each deme is approximated as a proportion m$\bar{x}$ (i.e. the average number of alleles of the first type in the metapopulation). This assumes that there is a sample pool of migrants from all demes which then migrate at random (so that migrants from a given deme can in principle return to the parental deme). While this is a reasonable assump‐tion for certain instances of a BMS life cycle (for instance, broadcast spawning marine organisms), it less realistic as an exact model for metapopulations under BSM.

In the BSM life cycle, it is adults (or at least some age class at which density dependence does not strongly act) that migrate. It is unrealistic to assume a "broadcast and random return" form of migration in this case, so that in actuality each deme only receives migrants from the D−1 other demes in the population (or from nearest neigh‐bors if there is spatial structure). However, treating migration in BSM as a pooled effect is a reasonable approximation when D is very large and/or allele frequencies do not greatly differ across demes. Furthermore, since in the BSM life cycle migration will be shown to be irrelevant to the effects of offspring variance, the results pertinent to the question of effective population size and fitness of a strategy are qualitatively the same regardless.

It should also be noted that an BSM life cycle is biologically equivalent to MBS and SMB since the events occur cyclically and in sequence (with the same reasoning applicable to BMS versus MSB, SBM). In the actual calculations, differences in starting point correspond to different stage of the life cycle at the census point. So while the formal expressions may differ for various choices of census point, it is rather obvious that the relations between terms do not change since the same processes are involved.

### ■ Birth→Selection→Migration Life Cycle

Consider first the life cycle where the sequence of events is birth→selection→migration (BSM). In the absence of spatial structure, every deme sends a proportion m of its population to each of the remaining D−1 demes. The change in the number of individuals of the first genotype x in the it deme due to migration is estimated as

$$x_i(t+1) = (1-m)x_i + m\bar{x},$$

where $\bar{x} = \frac{1}{D} \sum_{j=1}^{D} x_j$.

In exact terms, this corresponds to the case where every deme contributes a fraction m to a common migrant pool that is then divided up between the D demes (so that some of the "migrants" return to the parental population). A more realistic model (particularly for a life cycle where adults migrate) is one where the migrants can only move to neighboring demes. This mode of migration is represented by

$$x_i(t+1) = (1-m)x_i + \frac{m}{D-1}\sum_{j \neq i} x_j$$

(i.e. deme i sends fraction m to all of the D−1 other demes and receives migrants from the other demes in same ratios). Describing migration in terms of metapopulation mean $\bar{x}$ is an inexact but reasonable estimate if there is minimal difference in allele frequency between demes or if the number of demes is large (see above).

The frequency of the first allele in the ith deme after reproduction, selection, and migration is

(18) $p_i(t+1) =$

$$m \frac{\bar{x}\,\mu_1 + \epsilon \sum_{i=1}^{x_1} z_1[i]}{(\bar{x}\,\mu_1 + (n-\bar{x})\mu_2 + \epsilon(\sum_{i=1}^{x_1} z_1[i] + \sum_{i=1}^{x_2} z_2[i]))} +$$

$$(1-m) \frac{(x_i\,\mu_1 + \epsilon \sum_{i=1}^{x_1} z_1[i])}{(x_i\,\mu_1 + (n-x_i)\mu_2 + \epsilon(\sum_{i=1}^{x_1} z_1[i] + \sum_{i=1}^{x_2} z_2[i]))}$$

Writing a series expansion in terms of the first and second powers of $\epsilon$, p(t+1) is approximated by

$$(19) \quad \frac{m\,\bar{x}\,\mu_1}{\bar{x}\,\mu_1 + n\,\mu_2 - \bar{x}\,\mu_2} + \frac{(1-m)\,x_i\,\mu_1}{x_i\,\mu_1 + n\,\mu_2 - x_i\,\mu_2} +$$

$$\epsilon\left(m\left(\frac{S[z_1,\bar{x}]}{\bar{x}\,\mu_1 + n\,\mu_2 - \bar{x}\,\mu_2} - \frac{\bar{x}\,(S[z_1,\bar{x}] + S[z_2, n-\bar{x}])\,\mu_1}{(\bar{x}\,\mu_1 + n\,\mu_2 - \bar{x}\,\mu_2)^2}\right) + \right.$$

$$(1-m)\left(\frac{S[z_1,x_i]}{x_i\,\mu_1 + n\,\mu_2 - x_i\,\mu_2} - \frac{(S[z_1,x_i] + S[z_2, n-x_i])\,x_i\,\mu_1}{(x_i\,\mu_1 + n\,\mu_2 - x_i\,\mu_2)^2}\right)\right) +$$

$$\epsilon^2\left(m\left(\frac{\bar{x}\,(S[z_1,\bar{x}] + S[z_2, n-\bar{x}])^2\,\mu_1}{(\bar{x}\,\mu_1 + n\,\mu_2 - \bar{x}\,\mu_2)^3} - \frac{S[z_1,\bar{x}]\,(S[z_1,\bar{x}] + S[z_2, n-\bar{x}])}{(\bar{x}\,\mu_1 + n\,\mu_2 - \bar{x}\,\mu_2)^2}\right) + \right.$$

$$(1-m)\left(\frac{(S[z_1,x_i] + S[z_2, n-x_i])^2\,x_i\,\mu_1}{(x_i\,\mu_1 + n\,\mu_2 - x_i\,\mu_2)^3} - \frac{S[z_1,x_i]\,(S[z_1,x_i] + S[z_2, n-x_i])}{(x_i\,\mu_1 + n\,\mu_2 - x_i\,\mu_2)^2}\right)\right)$$

where $S[z,x] = \sum_{i=1}^{x} z[i]$.

Collecting terms with the same coefficients and applying the relationships $\mathrm{cov}(z_1[i], z_1[j])=0$, $\mathrm{cov}(z_1[i], z_2[j])$, and $\left(\sum_{i=1}^{x} z[i]\right)^2 = x\sigma^2/\epsilon^2$, together with the substitutions $x_1 = np$, $x_2 = n(1-p)$, $\bar{x} = n\bar{p}$, the final form for $M(p_i)$, the change in allele frequency across a generation, is the difference between the expectation of (19) and $p_i$.

$$(20) \quad M(p_i) = \left(\frac{m\,\bar{p}\,\mu_1}{\bar{p}\,\mu_1 + (1-\bar{p})\,\mu_2} + \frac{(1-m)\,p_i}{p_i\,\mu_1 + (1-p_i)\,\mu_2} - p_i\right) +$$

$$\left(\frac{m\,(1-\bar{p})\,\bar{p}\,\mu_1\,(\mu_2\,\sigma_1^2 + \mu_1\,\sigma_2^2)}{n\,(\bar{p}\,\mu_1 + (1-\bar{p})\,\mu_2)^3} + \frac{(1-m)\,p_i\,(1-p_i)\,(\mu_2\,\sigma_1^2 + \mu_1\,\sigma_2^2)}{n\,(p_i\,\mu_1 + (1-p_i)\,\mu_2)^3}\right)$$

Comparing (20) with the single deme equations (7), it can be seen that the terms in the first set of square brackets represent the "deterministic" change in allele frequency due to differences in allele frequency across demes, and, as in the single deme case, directional selection due to mean differences in fitness between the two alleles. The terms in the second set of square brackets represent selection on variance, which is largely independent of migration in this life cycle except for differences in relative frequency. When $p_i = \bar{p}$ (allele frequencies equal in all demes), the above expression reduces to an equation that is identical to (7–8), which is independent of the migration rate m because every $\bar{p} = p_i$ term with a factor of m has a complement $p_i$ factor of (1−m),

$$(21) \quad M(p_i) = \frac{p_i(1-p_i)(\mu_1 - \mu_2)}{p_i \mu_1 + (1-p_i)\mu_2} + \frac{p_i(1-p_i)(\mu_2 \sigma_1^2 + \mu_1 \sigma_2^2)}{n(p_i \mu_1 + (1-p_i)\mu_2)^3}.$$

Consequently, the selection dynamics defined by M(p) in a metapopulation of D demes of size n under the BSM life cycle are identical to that within a single deme of size n, regardless of migration rate. Migration only contributes to M(p) due to frequency differences between demes, not because of the accumulation of variance across demes. This can be seen from the fact that in (21) the variance terms scale inversely with n (as for a single deme) independent of migration rate and the total number of demes, so that the effective fitness of a strategy remains $\mu - \frac{\sigma^2}{n}$. Even when there are differences in allele frequencies across demes, the selection on variance is still of the order $\frac{\sigma^2}{n}$, with the only difference being that of an allelic (additive) variance factor of $p_i(1-p_i)\frac{\sigma^2}{n}$ versus the migration weighted term $(m\bar{p}(1-\bar{p}) + (1-m)p_i(1-p_i))\frac{\sigma^2}{n}$.

That migration only contributes to change in allele frequency when $p_i \neq \bar{p}$ can be seen in substituting the limiting cases of m=0 (no migration) and m=1 (complete mixing) into (20), which give, respectively,

$$(22.a) \quad \frac{p_i(1-p_i)(\mu_1-\mu_2)}{p_i\mu_1+(1-p_i)\mu_2} + \frac{p_i(1-p_i)(\mu_2\sigma_1^2+\mu_1\sigma_2^2)}{n(p_i\mu_1+(1-p_i)\mu_2)^3}$$

$$(22.b) \quad \frac{\bar{p}(1-\bar{p})(\mu_1-\mu_2)}{\bar{p}\mu_1+(1-\bar{p})\mu_2} + \frac{(1-\bar{p})\bar{p}\mu_1(\mu_2\sigma_1^2+\mu_1\sigma_2^2)}{n(\bar{p}\mu_1+(1-\bar{p})\mu_2)^3}$$

which differ only in the relative roles of $p_i$, $\bar{p}$ and not in the value of the denominator term associated with $\sigma^2$. Thus, selection can still act on the variance in offspring number even in populations that appear to be very large.

### ■ Birth→Migration→Selection Life Cycle

As was noted in the introduction, if reproduction in every deme is followed by migration prior to selective culling, the evolutionary dynamics are quite different from the BSM life cycle, because the offspring sample variance will be reduced by contributions from every deme. It is expected that this will be reflected in the form of M($p_i$) for the BMS life cycle, since the normalization of allele counts (representing soft selection) is done over absolute frequencies across as well as within demes.

Following the same logic as in the last derivations, the frequency of the first allele at the end of the BMS life cycle is

$$(23) \quad p_i(t+1) = \\ \left(m\left(\bar{x}\mu_1+\frac{\in S[z_1,D\bar{x}]}{D}\right)+(1-m)(x_i\mu_1+\in S[z1,x_i])\right) \Big/ \\ \left(m\left(\bar{x}\mu_1+\frac{\in S[z_2,D\bar{x}]}{D}\right)+(1-m)(x_i\mu_1+\in S[z1,x_i])+ \right. \\ \left. m\left((n-\bar{x})+\mu_2\frac{\in S[z_2,D(n-\bar{x})]}{D}\right)+ \right. \\ \left. (1-m)((n-x_i)\mu_2+\in S[z_2,n-x_i])\right)$$

Here migration is also assumed to occur via the "mixing" used in the previous derivations, i.e. all demes contribute a certain fraction of their offspring to a migrant pool which then distributes at random across all demes (allowing for return migration and Here migration is also assumed to occur via the "mixing" used in the previous derivations, i.e. all demes contribute a certain fraction of their offspring to a migrant pool which then distributes at random across all demes (allowing for return migration and thus the characterization of the migrant pool as an average $\bar{x}$). It will be shown that for this life cycle, the number of alleles in the entire metapopulation $D\bar{x}$ contributes to the variance component in every deme.

To calculate the change in $p_i$ in terms of offspring variance, (23) is expanded as a power series up to second order in terms of $\epsilon$,

(24)

$$\frac{m\bar{x}\mu_1 + x_i\mu_1 - mx_i\mu_1}{m\bar{x}\mu_1 + (1-m)x_i\mu_1 + m(n-\bar{x})\mu_2 + (1-m)(n-x_i)\mu_2} +$$

$$\epsilon\left(\left(\frac{mS[z_1, D\bar{x}]}{D} + (1-m)S[z_1, x_i]\right) \Big/ (m\bar{x}\mu_1 + (1-m)x_i\mu_1 + m(n-\bar{x})\mu_2 + (1-m)(n-x_i)\mu_2) - \left(\left(\frac{mS[z_1, D\bar{x}]}{D} + S[z_1, x_i] - mS[z_1, x_i] + \frac{mS[z_2, D(n-\bar{x})]}{D} + S[z_2, n-x_i] - mS[z_2, n-x_i]\right)(m\bar{x}\mu_1 + (1-m)x_i\mu_1)\right) \Big/ (m\bar{x}\mu_1 + (1-m)x_i\mu_1 + m(n-\bar{x})\mu_2 + (1-m)(n-x_i)\mu_2)^2\right) +$$

$$\epsilon^2\left(\left(\left(\frac{mS[z_1, D\bar{x}]}{D} + S[z_1, x_i] - mS[z_1, x_i] + \frac{mS[z_2, D(n-\bar{x})]}{D} + S[z_2, n-x_i] - mS[z_2, n-x_i]\right)^2 (m\bar{x}\mu_1 + (1-m)x_i\mu_1)\right) \Big/ (m\bar{x}\mu_1 + (1-m)x_i\mu_1 + m(n-\bar{x})\mu_2 + (1-m)(n-x_i)\mu_2)^3 - \left(\left(\frac{mS[z_1, D\bar{x}]}{D} + S[z_1, x_i] - mS[z_1, x_i]\right)\right)$$

$$\left(\frac{m\,S[z_1, D\bar{x}]}{D} + S[z_1, x_i] - \right.$$
$$m\,S[z_1, x_i] + \frac{m\,S[z_2, D(n-\bar{x})]}{D} +$$
$$\left.S[z_2, n-x_i] - m\,S[z_2, n-x_i]\right)\Big/$$
$$(m\,\bar{x}\,\mu_1 + (1-m)\,x_i\,\mu_1 + m\,(n-\bar{x})\,\mu_2 +$$
$$(1-m)(n-x_i)\,\mu_2)^2\Big)$$

M($p_i$) is the difference between the expectation of the above expression and $p_i$. Applying the variance and covariance relations and substituting np=$x_1$, n(1−p)=$x_2$, n$\bar{p}$=$\bar{x}$, the expression becomes:

$$(25) \quad M(p_i) =$$
$$\frac{m\,\bar{p}\,\mu_1 + (1-m)\,p_i\,\mu_1}{m\,\bar{p}\,\mu_1 + (1-m)\,p_i\,\mu_1 + m\,(1-\bar{p})\,\mu_2 + (1-m)(1-p_i)\,\mu_2} -$$
$$\left(\frac{m^2\,\bar{p}\,\sigma_1^2}{D} - (1-m)^2\,p_i\,\sigma_1^2\right)\Big/n\,(m\,\bar{p}\,\mu_1 + (1-m)\,p_i\,\mu_1 +$$
$$m\,(1-\bar{p})\,\mu_2 + (1-m)(1-p_i)\,\mu_2)^2 +$$
$$\left((m\,\bar{p}\,\mu_1 + (1-m)\,p_i\,\mu_1)\left(\frac{m^2\,n\,\bar{p}\,\sigma_1^2}{D} + (1-m)^2\,n\,p_i\,\sigma_1^2 +\right.\right.$$
$$\left.\left.\frac{m^2(1-\bar{p})\,\sigma_2^2}{D} + (1-m)^2(1-p_i)\,\sigma_2^2\right)\right)\Big/n$$
$$(m\,\bar{p}\,\mu_1 + (1-m)\,p_i\,\mu_1 + m\,(1-\bar{p})\,\mu_2 +$$
$$(1-m)(1-p_i)\,\mu_2)^3 - p_i$$

In the case where allele frequencies are equal ($p_i=\bar{p}$) across all demes, the mean change in frequency simplifies to

$$(26) \quad M(p_i) = \frac{p_i(1-p_i)(\mu_1 - \mu_2)}{p_i\,\mu_1 + (1-p_i)\,\mu_2} -$$
$$\left((1-m)^2\,p_i\,\sigma_1^2 + \frac{m^2\,p_i\,\sigma_1^2}{D}\right)\Big/n\,(p_i\,\mu_1 + (1-p_i)\,\mu_2)^2 +$$
$$\left(p_i\,\mu_1\left((1-m)^2\,p_i\,\sigma_1^2 + \frac{m^2\,p_i\,\sigma_1^2}{D} + (1-m)^2(1-p_i)\,\sigma_2^2 +\right.\right.$$
$$\left.\left.\frac{m^2(1-p_i)\,\sigma_2^2}{D}\right)\right)\Big/n\,(p_i\,\mu_1 + (1-p_i)\,\mu_2)^3$$

It is apparent that the above expression is not equivalent to M(p) for a single deme (Eq. 7). Even when allele frequencies are equal across demes, migration has the indirect effect of reducing variance by sampling from the pool of offspring throughout the metap−

It is apparent that the above expression is not equivalent to M(p) for a single deme (Eq. 7). Even when allele frequencies are equal across demes, migration has the indirect effect of reducing variance by sampling from the pool of offspring throughout the metapopulation.

In the limiting cases of m=0 and m=1, the respective values of $M(p_i)$ are

(27.a)
$$\frac{(1-p_i) p_i (\mu_1 - \mu_2)}{p_i \mu_1 + (1-p_i) \mu_2} - \frac{p_i \sigma_1^2}{n (p_i \mu_1 + (1-p_i) \mu_2)^2} + \frac{p_i \mu_1 (p_i \sigma_1^2 + (1-p_i) \sigma_2^2)}{n (p_i \mu_1 + (1-p_i) \mu_2)^3}$$

(27.b)
$$\frac{(1-\bar{p}) \bar{p} (\mu_1 - \mu_2)}{\bar{p} \mu_1 + (1-\bar{p}) \mu_2} - \frac{\bar{p} \sigma_1^2}{D n (\bar{p} \mu_1 + (1-\bar{p}) \mu_2)^2} + \frac{\bar{p} \mu_1 (\bar{p} \sigma_1^2 + (1-\bar{p}) \sigma_2^2)}{Dn (\bar{p} \mu_1 + (1-\bar{p}) \mu_2)^3}$$

For a BMS life cycle then, migration rate determines the effective population size "seen" by selection acting on offspring variance. When m=0, each deme is independent of the others, so that M(p) is the same as (7) for a single deme (i.e. n in the denominator of the variance terms). When m=1 (corresponding to a case of complete mixing), the denominator is nD, the total metapopulation size. In contrast with BSM, even when $p_i = \bar{p}$, it can be seen that (27.a) and (27.b) are obviously not equivalent: one has a variance component that scales with deme size n, the other with metapopulation size.

These results are largely consistent with the numerical findings in Shpak (2005) for the limiting cases. On heuristic grounds, it was argued that the effective population size (denominator of the variance terms in M(p)) would scale as (1−m)n+mDn, as a function of migration rate. The form of equation 25 suggests that the relationship is actually more complicated for intermediate values of m. The variance contributions to M(p) actually scale as functions of the squares of migration rates, with n in the denominator for "residents" and nD for the "migrants,"

$$(28) \quad \frac{(1-m)^2 \, p_i \, \sigma_1^2}{n} + \frac{m^2 \, \bar{p} \, \sigma_1^2}{nD}$$

if this is compared to $\frac{p\sigma^2}{n}$ for a single deme, then it can be argued that migration in the BMS model induces an effective population size from the standpoint of selection for variance, so that

$$(29) \quad \frac{p\sigma^2}{n_e} = \frac{(1-m)^2 \, p_i \, \sigma_1^2}{n} + \frac{m^2 \, \bar{p} \, \sigma_1^2}{nD}$$

Ignoring the effects of frequency differences between demes and setting $\bar{p}=p_i=p$,

$$(30) \quad n_e \approx \frac{nD}{D(1-m)^2 + m^2},$$

which increases towards nD as m approaches unity and decreases towards n as m approaches 0. This yields the prediction that if $\mu_1 < \mu_2$, $\sigma_1^2 < \sigma_2^2$ and n is small enough so that $\mu_2 - \frac{\sigma_2^2}{n} < \mu_1 - \frac{\sigma_1^2}{n}$, for sufficiently many demes and a high enough migration rate there will be a critical value m such that $n_e$ in (30) satisfies Eq. 17 (i.e. the strategy with higher effective fitness at zero or low migration is disfavored at high migration rates because the selection against variance is less pronounced in a higher effective population size). We explore this theme further in the discussion.

The sign of M(p) determines whether an allele is expected to increase or decrease in frequency given a set of parameters (describing mean and variance in offspring number, population size, and migration rate). Since there is no frequency dependence, M(p) has the same sign for any value of p, so that if M(p)>0, the first allele is favored by selection, while M(p)<0 implies that the second allele is favored. The value and sign of M(p) in itself does not provide sufficient information to predict the probabil-ity of loss or fixation, however, since there is a stochastic contribution to the dynamics represented by the diffusion term V(p). Rather, consideration of V(p) allows us to infer relative fixation probabilities and qualitatively describe evolutionary trajectories (Proulx and Day 2001, Nowak 2004, Wild and Taylor 2004).

The derivations of V($p_i$) for a metapopulation are relegated to the Appendix. It should be noted that a complete description of fixation probabilities in a finite population would also incorporate the contribution of genetic drift proper to V(p) (e.g. Proulx 2000), which would add binomial sampling probabilities to the already involved equations for offspring variance. In the appendix, V(p) is only calculated for the stochastic contribution of clutch size variance.

## Discussion

There are a several potentially important consequences of selection on variance of offspring production in metapopulations. That the effective fitness of a strategy where there is variance in clutch size depends on population size was established by Gillespie (1974), so it stands to reason that population processes that lead to differences between census size and "effective" population size (such as density fluctuations, differences in frequency between the sexes, (see Proulx 2000) can lead to differences in the selective advantage of a strategy with a given mean and variance.

Migration and population structure are known to cause a discrepancy between census and effective population size from the standpoint of genetic drift (e.g. Whitlock and Barton 1997), which might suggest that other sources of variance would be influenced by subdivision. It is interesting that these effects turn out to depend on the sequence of events in an organism's life cycle, i.e. whether reproduction and selection take place within demes prior to migration or whether migration occurs after reproduc-tion but before selective culling. In the BSM life cycle, the effective population size of a deme is essentially equal to its census size, while for BMS, the effective size of a deme (and for that matter, the entire metapopulation as an average of behavior across demes) ranges from the census size of a deme to the census size of a metapopulation.

What this suggests is that the influence of offspring variance on the performance of a given strategy cannot necessarily be ignored even in large populations. Tradition–ally, the significance of equations (8–9) was considered by many to be of only aca–demic interest on the grounds that most biological populations were large enough for sample variance to play a minor role. What the results in this paper show is that even in a large metapopulation with extensive mixing, if the life cycle BSM and the population consists of many small demes, the effect of offspring variance is essentially the same as it would be for a single small deme. Consequently, the impact of offspring variance must be assessed on a case by case basis, depending on life cycle and population struc–ture as well as the fitness and variance parameters.

In other words, if there is a trade–off between producing a high mean number of offspring and reducing variance (as in the semelparous vs. iteroparous regimes men–tioned in the introduction), the outcome of selection will depend not only on the census number in the population, but on life cycle and migration rate. Using the semelparity and iteroparity examples, an organism that reproduces once and produces an average of on clutch $k_1=1$ of $\omega_1=10$ offspring that survive (as a whole clutch) with a probability of $\pi=0.1$ and fail with a probability of $1-\pi=0.9$ has a mean fitness $\mu_1=k_1\omega_1\pi=1$ and vari–ance $\sigma_1^2=k_1\omega_1^2\pi(1-\pi)=9$. If it competes against an iteroparous strategy that produces a $k_2=10$ clutches of $\omega_2=1$ single offspring (with the probability of surviving $\pi=0.1$), the parameters are $\mu_2=0.9$ and $\sigma_2^2=0.81$.

Equation (17) predicts that for a population size n<81.9 the iteroparous strategy will be favored in spite of having a lower arithmetic mean. If the selection takes place in the context of a metapopulation with D=10 demes with n=50 individuals per deme, the iteroparous strategy will always be favored in BSM life cycle regardless of migration rate.

In a BMS life cycle, there will be a critical value of migration rate at which the higher mean semelparous strategy starts to be favored. This critical value can be approxi–mated as the value of m that gives $n_e = 81.9$ in Eq. (30), with n=50 and D=10. Solving the quadratic equation for m, the root less then unity is m=0.2218. This corresponds reasonably well with the individual based simulations in Shpak (2005), where the proba–bility of fixing the iteroparous strategy (with initial frequency p=0.5) was near 50% when there were between 1 and 2 migrants exchanged between any pair of demes (corresponding to effective neutrality for some value of m between 0.2 and 0.4). The simulations include effects of genetic drift, but (30) still gives a much better estimate of the critical migration rate than the linear estimate in Shpak 2005.

These results predict broad trends in the evolution of life histories and reproduc–tive strategies in various organisms. Since for the same mean value of offspring an iterop-arous strategy produces a lower variance in surviving progeny than a semelparous strat–egy, it is predicted that semelparity should be less common in organisms with small population sizes, or in highly structured populations when the life cycle is of the BSM type. For a large population or a metapopulation of organisms with a BMS life cycle, the penalty for high variance is lower and semelparity should be more common, particularly if the semelparous strategy can produce a higher mean number of offspring.

At a practical level, this means that one would expect semelparity to be more common among organisms where the most incoming migrants to any deme are juveniles and the most important selection takes place after migration. This is the case with broad–cast spawning marine invertebrates, many of which have large, widely distributed metap–opulations that exchange migrants via planktonic eggs and larvae, and with plants that disperse seeds over long distances. Semelparity would probably be more prevalent in such organisms than in (for example) birds or large mammals, where most of the migra–tion between demes is by adults that have already been subjected to an entire life of selection.

The trade–offs involved in determining reproductive strategy are often more complicated than a balance between mean and variance (for example, a balance between adult and juvenile mortality, as discussed by Charnov and Schaffer 1973), but the effects of offspring variance on the fitness of a genotype are potentially strong enough for this to be an important factor in many animal and plant populations.

## Appendix: The Diffusion Coefficient in Metapopulation Model

The derivation of V(p) in a single deme is shown in Equations 11–14. The same approxi-mations and assumptions about migration used to derive E[p(t+1)] in a metapopulation under different life cycles is used to calculate the second moments:

### ■ Birth→Selection→Migration Life Cycle

$$(A.1) \quad p(t+1)^2 = \left( \frac{m (\bar{x} \mu_1 + \epsilon \, s[z_1, \bar{x}])}{\epsilon (s[z_1, \bar{x}] + s[z_2, n - \bar{x}]) + \bar{x} \mu_1 + (n - \bar{x}) \mu_2} + \frac{(1 - m) (x_i \mu_1 + \epsilon \, s[z_1, x_i])}{\epsilon (s[z_1, x_i] + s[z_2, n - x_i]) + x_i \mu_1 + (n - x_i) \mu_2} \right)^2$$

writing a Taylor expansion in terms of $\epsilon$,

$$(A.2) \quad \left( \frac{m \bar{x} \mu_1}{\bar{x} \mu_1 + (n - \bar{x}) \mu_2} + \frac{(1 - m) x_i \mu_1}{x_i \mu_1 + (n - x_i) \mu_2} \right)^2 +$$

$$2 \epsilon \left( \frac{m \bar{x} \mu_1}{\bar{x} \mu_1 + (n - \bar{x}) \mu_2} + \frac{(1 - m) x_i \mu_1}{x_i \mu_1 + (n - x_i) \mu_2} \right)$$

$$\left( m \left( \frac{S[z_1, \bar{x}]}{\bar{x} \mu_1 + (n - \bar{x}) \mu_2} - \frac{\bar{x} (S[z_1, \bar{x}] + S[z_2, n - \bar{x}]) \mu_1}{(\bar{x} \mu_1 + (n - \bar{x}) \mu_2)^2} \right) + \right.$$

$$(1 - m) \left( \frac{S[z_1, x_i]}{x_i \mu_1 + (n - x_i) \mu_2} - \right.$$

$$\left. \left. \frac{(S[z_1, x_i] + S[z_2, n - x_i]) x_i \mu_1}{(x_i \mu_1 + (n - x_i) \mu_2)^2} \right) \right) +$$

$$\epsilon^2 \left( 2 \left( \frac{m \bar{x} \mu_1}{\bar{x} \mu_1 + (n - \bar{x}) \mu_2} + \frac{(1 - m) x_i \mu_1}{x_i \mu_1 + (n - x_i) \mu_2} \right) \right.$$

$$\left( m \left( \frac{\bar{x} (S[z_1, \bar{x}] + S[z_2, n - \bar{x}])^2 \mu_1}{(\bar{x} \mu_1 + (n - \bar{x}) \mu_2)^3} - \right. \right.$$

$$\left. \frac{S[z_1, \bar{x}] (S[z_1, \bar{x}] + S[z_2, n - \bar{x}])}{(\bar{x} \mu_1 + (n - \bar{x}) \mu_2)^2} \right) +$$

$$(1 - m) \left( \frac{(S[z_1, x_i] + S[z_2, n - x_i])^2 x_i \mu_1}{(x_i \mu_1 + (n - x_i) \mu_2)^3} - \right.$$

$$\left. \left. \frac{S[z_1, x_i] (S[z_1, x_i] + S[z_2, n - x_i])}{(x_i \mu_1 + (n - x_i) \mu_2)^2} \right) \right) +$$

$$\left( m \left( \frac{S[z_1, \bar{x}]}{\bar{x} \mu_1 + (n - \bar{x}) \mu_2} - \right. \right.$$

$$\left. \frac{\bar{x} (S[z_1, \bar{x}] + S[z_2, n - \bar{x}]) \mu_1}{(\bar{x} \mu_1 + (n - \bar{x}) \mu_2)^2} \right) +$$

$$(1 - m) \left( \frac{S[z_1, x_i]}{x_i \mu_1 + (n - x_i) \mu_2} - \right.$$

$$\left. \left. \left. \frac{(S[z_1, x_i] + S[z_2, n - x_i]) x_i \mu_1}{(x_i \mu_1 + (n - x_i) \mu_2)^2} \right) \right)^2 \right)$$

applying the variance and covariance relations on the sums of z (reducing the expression to functions of the mean and variance values) and substituting $np_i$, $n\bar{p}$ for $x_i$ and $\bar{x}$, the expectation value is

$$(A.3) \quad E[p(t+1)^2] = \left( \frac{m \bar{p} \mu_1}{\bar{p} \mu_1 + (1-\bar{p}) \mu_2} + \frac{(1-m) p_i \mu_1}{p_i \mu_1 + (1-p_i) \mu_2} \right)^2 +$$

$$\frac{m^2 (1-\bar{p}) \bar{p} ((1-\bar{p}) \mu_2^2 \sigma_1^2 - \bar{p} \mu_1^2 \sigma_2^2)}{n (\bar{p} (\mu_1 - \mu_2) + \mu_2)^4} +$$

$$\frac{(1-m)^2 (1-p_i) p_i ((1-p_i) \mu_2^2 \sigma_1^2 - p_i \mu_1^2 \sigma_2^2)}{n (p_i (\mu_1 - \mu_2) + \mu_2)^4} +$$

$$2 \left( \frac{m \bar{p} \mu_1}{\bar{p} \mu_1 + (1-\bar{p}) \mu_2} + \frac{(1-m) p_i \mu_1}{p_i \mu_1 + (1-p_i) \mu_2} \right)$$

$$\left( m \frac{(1-\bar{p}) \bar{p} \mu_1 (\mu_2 \sigma_1^2 + \mu_1 \sigma_2^2)}{n (\bar{p} \mu_1 + (1-\bar{p}) \mu_2)^3} + \right.$$

$$\left. (1-m) \frac{p_i (1-p_i) (\mu_2 \sigma_1^2 + \mu_1 \sigma_2^2)}{n (p_i \mu_1 + (1-p_i) \mu_2)^3} \right)$$

The expression for $V(p_i)$ is simply the above quantity plus $(p_i^2 - 2 p_i \, E[p_i(t+1)])$, where $E[p_i(t+1)]$ consists of the terms in Equations 19–20.

As was the case for the first moments, there is no contribution of offspring variance to $V(p_i)$ in the BSM life cycle except for the weighted difference of allele frequencies between the metapopulation mean and the census deme. In the absence of allele frequency differences, $V(p_i)$ reduces to an equation identical to (14) for a single deme, as did the first moment.

### ■ Birth→Migration→Selection Life Cycle

The second moment of $p_i(t+1)$ is calculated from the expectation of:

$$(A.4) \quad p^2(t+1) = \left( m \left( \frac{\in s[z_1, D \bar{x}]}{D} + \bar{x} \mu_1 \right) + (1-m) (\in s[z_1, x_i] + x_i \mu_1) \right)^2 \Big/$$

$$\left(m\left(\frac{\epsilon\, S[z_1, D\bar{x}]}{D} + \bar{x}\,\mu_1\right) + (1-m)\,(\epsilon\, S[z_1, x_i] + x_i\,\mu_1) + \right.$$
$$m\left(\frac{\epsilon\, S[z_2, D\,(n-\bar{x})]}{D} + (n-\bar{x})\,\mu_2\right) +$$
$$\left.(1-m)\,(\epsilon\, S[z_2, n-x_i] + (n-x_i)\,\mu_2)\right)^2$$

Up to the quadratic term in a series expansion about $\epsilon=0$, the above expression is:

$$(A.5) \quad (m\,\bar{x}\,\mu_1 + (1-m)\,x_i\,\mu_1)^2 \,/\,$$
$$(m\,\bar{x}\,\mu_1 + (1-m)\,x_i\,\mu_1 + m\,(n-\bar{x})\,\mu_2 +$$
$$(1-m)\,(n-x_i)\,\mu_2)^2 +$$
$$\epsilon\left(\left(2\left(\frac{m\,S[z_1, D\,\bar{x}]}{D} + S[z_1, x_i] - m\,S[z_1, x_i]\right)\right.\right.$$
$$\left.(m\,\bar{x}\,\mu_1 + (1-m)\,x_i\,\mu_1)\right)\,/\,(m\,\bar{x}\,\mu_1 + (1-m)\,x_i\,\mu_1 +$$
$$m\,(n-\bar{x})\,\mu_2 + (1-m)\,(n-x_i)\,\mu_2)^2 -$$
$$\left(2\left(\frac{m\,S[z_1, D\,\bar{x}]}{D} + S[z_1, x_i] - m\,S[z_1, x_i] + \right.\right.$$
$$\frac{m\,S[z_2, D\,(n-\bar{x})]}{D} + S[z_2, n-x_i] -$$
$$\left.\left.m\,S[z_2, n-x_i]\right)\,(m\,\bar{x}\,\mu_1 + (1-m)\,x_i\,\mu_1)^2\right)\,/\,$$
$$\left.(m\,\bar{x}\,\mu_1 + (1-m)\,x_i\,\mu_1 + m\,(n-\bar{x})\,\mu_2 +\right.$$
$$\left.(1-m)\,(n-x_i)\,\mu_2)^3\right) +$$
$$\epsilon^2\left(\left(3\left(\frac{m\,S[z_1, D\,\bar{x}]}{D} + S[z_1, x_i] - m\,S[z_1, x_i] + \right.\right.\right.$$
$$\frac{m\,S[z_2, D\,(n-\bar{x})]}{D} +$$
$$\left.S[z_2, n-x_i] - m\,S[z_2, n-x_i]\right)^2$$
$$\left.(m\,\bar{x}\,\mu_1 + (1-m)\,x_i\,\mu_1)^2\right)\,/\,(m\,\bar{x}\,\mu_1 + (1-m)\,x_i\,\mu_1 +$$
$$m\,(n-\bar{x})\,\mu_2 + (1-m)\,(n-x_i)\,\mu_2)^4 -$$
$$\left(4\left(\frac{m\,S[z_1, D\,\bar{x}]}{D} + S[z_1, x_i] - m\,S[z_1, x_i]\right)\right.$$
$$\left(\frac{m\,S[z_1, D\,\bar{x}]}{D} + S[z_1, x_i] - m\,S[z_1, x_i] + \right.$$

$$\frac{m\,S[z_2,\,D\,(n-\bar{x})]}{D} + S[z_2,\,n-x_i] -$$

$$m\,S[z_2,\,n-x_i]\bigg)\,(m\,\bar{x}\,\mu_1 + (1-m)\,x_i\,\mu_1)\bigg)\bigg/$$

$$(m\,\bar{x}\,\mu_1 + (1-m)\,x_i\,\mu_1 + m\,(n-\bar{x})\,\mu_2 +$$

$$(1-m)\,(n-x_i)\,\mu_2)^3 +$$

$$\left(\frac{m\,S[z_1,\,D\,\bar{x}]}{D} + S[z_1,\,x_i] - m\,S[z_1,\,x_i]\right)^2\bigg/$$

$$(m\,\bar{x}\,\mu_1 + (1-m)\,x_i\,\mu_1 +$$

$$m\,(n-\bar{x})\,\mu_2 + (1-m)\,(n-x_i)\,\mu_2)^2\bigg)$$

Substituting mean and variance values for S[z,x], the expectation is

$$(A.6) \quad E[p^2(t+1)] =$$
$$(m\,\bar{p}\,\mu_1 + (1-m)\,p_i\,\mu_1)^2 \,/\, (m\,\bar{p}\,\mu_1 + (1-m)\,p_i\,\mu_1 +$$
$$m\,(1-\bar{p})\,\mu_2 + (1-m)\,(1-p_i)\,\mu_2)^2 -$$
$$\left(4\,(m\,\bar{p}\,\mu_1 + (1-m)\,p_i\,\mu_1)\left(\frac{m^2\,\bar{p}\,\sigma_1^2}{D} + (1-m)^2\,p_i\,\sigma_1^2\right)\right)\bigg/n$$
$$(m\,\bar{p}\,\mu_1 + (1-m)\,p_i\,\mu_1 +$$
$$m\,(1-\bar{p})\,\mu_2 + (1-m)\,(1-p_i)\,\mu_2)^3 +$$
$$\left(\frac{m^2\,\bar{p}\,\sigma_1^2}{D} + (1-m)^2\,p_i\,\sigma_1^2\right)\bigg/n\,(m\,\bar{p}\,\mu_1 + (1-m)\,p_i\,\mu_1 +$$
$$m\,(1-\bar{p})\,\mu_2 + (1-m)\,(1-p_i)\,\mu_2)^2 +$$
$$\left(3\,(m\,\bar{p}\,\mu_1 + (1-m)\,p_i\,\mu_1)^2\left(\frac{m^2\,\bar{p}\,\sigma_1^2}{D} + (1-m)^2\,p_i\,\sigma_1^2 +\right.\right.$$
$$\left.\left.\frac{m^2\,(1-\bar{p})\,\sigma_2^2}{D} + D\,(1-m)^2\,(1-p_i)\,\sigma_2^2\right)\right)\bigg/n$$
$$(m\,\bar{p}\,\mu_1 + (1-m)\,p_i\,\mu_1 + m\,(1-\bar{p})\,\mu_2 + (1-m)\,(1-p_i)\,\mu_2)^4$$

When the allele frequency in the ith deme equals that of the population mean, this reduces to

$$(A.7) \quad ((1-m)\,p_i\,\mu_1 + m\,p_i\,\mu_1)^2 \,/\, (p_i\,\mu_1 + (1-p_i)\,\mu_2)^2 -$$
$$\left(4\,((1-m)\,p_i\,\mu_1 + m\,p_i\,\mu_1)\left((1-m)^2\,p_i\,\sigma_1^2 + \frac{m^2\,p_i\,\sigma_1^2}{D}\right)\right)\bigg/n$$
$$(p_i\,\mu_1 + (1-p_i)\,\mu_2)^3 +$$

$$\left((1-m)^2 p_i \sigma_1^2 + \frac{m^2 p_i \sigma_1^2}{D}\right) \Big/ n (p_i \mu_1 + (1-p_i) \mu_2)^2 +$$

$$\left(3 p_i \mu_1^2 \left((1-m)^2 p_i \sigma_1^2 + \frac{m^2 p_i \sigma_1^2}{D} + D(1-m)^2 (1-p_i) \sigma_2^2 + \frac{m^2 (1-p_i) \sigma_2^2}{D}\right)\right) \Big/ n (p_i \mu_1 + (1-p_i) \mu_2)^4$$

which, as with the first moment, remains dependent on m and D in the BSM life even with equal allele frequencies.

V($p_i$) is calculated by adding the quantity ($p_i^2 - 2p_i E[p_i(t+1)]$) to A.6 or A.7, where E[p(t+1)] is given in the first terms of Eq. (25).

Since there are closed form solutions for V($p_i$) in a metapopulation, one can in principle calculate probabilities of fixation and loss from the Kolmogorov backward equation (10) using the integration in (16) for various parameters.

## Acknowledgements

Max Shpak wishes to thank Sergey Gavrilets for his advice and support (via NSF grant DEB−0111613 and NIH grant GM56693) during the writing of this paper, as well as David Waxman for commenting on an earlier incarnation of the derivations.